\documentclass[aps,pre,preprint,superscriptaddress,showpacs]{revtex4}
\usepackage{graphicx,amssymb,array,verbatim,color}
\newcolumntype{C}[1]{>{\centering\arraybackslash}p{#1}}

\def\eq#1{Eq.~(\ref{#1})}
\def\fig#1{Fig.~\ref{#1}}

\usepackage{color}

\begin{document}

\title{Contractile network models for adherent cells}
\author{P. Guthardt Torres}
\affiliation{Heidelberg University, Institute for Theoretical
Physics, Philosophenweg 19, 69120 Heidelberg, Germany}
\affiliation{Heidelberg University, Bioquant, Im 
Neuenheimer Feld 267, 69120 Heidelberg, Germany}
\author{I.B. Bischofs}
\email[]{I.Bischofs@zmbh.uni-heidelberg.de}
\affiliation{Heidelberg University, ZMBH, Im Neuenheimer Feld 282,
69120 Heidelberg, Germany}
\affiliation{Heidelberg University, Bioquant, Im Neuenheimer Feld 267,
69120 Heidelberg, Germany}
\author{U.S. Schwarz}
\email[]{Ulrich.Schwarz@bioquant.uni-heidelberg.de}
\affiliation{Heidelberg University, Institute for Theoretical
Physics, Philosophenweg 19, 69120 Heidelberg, Germany}
\affiliation{Heidelberg University, Bioquant, Im Neuenheimer Feld 267,
69120 Heidelberg, Germany}
\date{\today}

\begin{abstract}
Cells sense the geometry and stiffness of their adhesive environment
by active contractility. For strong adhesion to flat substrates,
two-dimensional contractile network models can be used to understand
how force is distributed throughout the cell. Here we compare the
shape and force distribution for different variants of such network
models.  In contrast to Hookean networks, cable networks reflect the
asymmetric response of biopolymers to tension versus compression. For
passive networks, contractility is modeled by a
reduced resting length of the mechanical links. In actively
contracting networks, a constant force couple is introduced into each
link in order to model contraction by molecular motors.  If combined
with fixed adhesion sites, all network models lead to invaginated cell
shapes, but only actively contracting cable networks lead to the circular arc
morphology typical for strongly adhering cells. In this case, shape
and force distribution are determined by local rather than global
determinants and thus are suited to endow the cell with a robust sense
of its environment. We also discuss non-linear and adaptive linker
mechanics as well as the relation to tissue shape.
\end{abstract}

\pacs{87.16.Ln, 87.10.-e, 87.17.Rt}

\maketitle

\section{Introduction}

During the last decade, it has been increasingly realized that
adherent cells actively explore and respond to the geometry and
stiffness of their adhesive environment, with dramatic consequences
for fundamental cellular processes such as survival, proliferation,
differentiation and migration~\cite{janmey09,geiger09}. Using chemical
inhibitors for both myosin motors and actin filaments, it has been
shown that active contractility of the actin cytoskeleton is an
indispensable requirement for the observed cellular response to the
physical properties of the environment. Although more and more details
are revealed regarding the molecular mechanisms underlying these
sensing
processes~\cite{geiger_adhesome1,geiger_adhesome2,hoffman_dynamic_2011},
what is still missing is a systems-level understanding of how cells
integrate the way force is generated, distributed, and sensed over the
whole cell. Therefore theoretical models are required which describe
how force is propagated through the actin cytoskeleton as a function
of environmental geometry and stiffness. Because the actin
cytoskeleton is an integral part of the sensing capabilities of cells,
one expects that its mechanical properties have evolved to support
these important cellular functions.

Although it is a standard procedure in experiments to block force
generation and propagation with chemical inhibitors or by RNA
interference, it is very difficult to measure how force is distributed
inside cells and between cells and their environment. Different
experimental approaches have been developed to meet this
challenge. Traction force microscopy allows to measure the forces
transmitted to the substrate~\cite{dembo99,butler,schwarz,tan_cells_2003,sabass}.
Recently this technique has been extended in such a way that also
cell-cell forces can be estimated from cell-matrix forces
\cite{liu_mechanical_2010,uss:maru11}. However, it is important to note that many forces
balance inside the cell and are not transmitted to the substrate, so
the forces existing inside cells might be much higher than appreciated
from traction force microscopy~\cite{bischofsprl09}. Laser cutting
allows to estimate forces from the mechanical relaxation after cutting
load-carrying elements like microtubules in the mitotic
spindle~\cite{grill2001,grill2003} or stress fibers in the actin
cytoskeleton~\cite{kumar,kumar2010,colombelli,waterman}. Laser
ablation can be used for subcellular analysis of cortical
tension~\cite{julichergrillnature}. However, these experiments only
probe local relaxation events of prominent cytoskeletal structures and
therefore might miss the global effects of more distributed and less
visible structures. Micromanipulation can be used to distort the
mechanical balance of the cell
globally~\cite{riveline,paul,heilspatz,merkel}, but the resulting
changes in force distribution can only be estimated indirectly from
its effects, e.g.\, growth of focal adhesions. To achieve a more
systematic understanding, these experimental approaches have to be
complemented by theoretical approaches.

Because the mechanical properties of cells, extracellular matrix and
tissue are strongly determined by filamentous networks of proteins
like actin, tubulin, lamin, spectrin or collagen,
mechanical networks are widely used theoretical models for cell and
tissue mechanics~\cite{b:howa01,boal}. One of the best studied cases
is the red blood cell, whose shape and mechanics has been studied with
network approaches in very large detail~\cite{hansen1,hansen2,
  boaldischer1,boaldischer2,limwortis,noguchigompper,sureshbpj,
  sureshpnas,fedosov}. Modern computer power permits to simulate each
of the roughly $10^5$ spectrin links separately and with molecular
detail, for example using the appropriately parametrized
force-extension curve of a semiflexible polymer~\cite{sureshbpj}. In
the limit of small extensions, these models usually reduce to Hookean
networks.

For adherent cells, the main structural determinant is the actin
cytoskeleton, whose mechanics differs in several important aspects
from the one of the spectrin network of red blood cells. In general,
the molecular structure of the actin network is much less defined.
Its most prominent feature in adhesion is strong contractility due to
activity of myosin II motors. This observation implies that the
mechanical links between the nodes of the network cannot be simple
actin filaments, but have to be bundles of actin filaments crosslinked
and tensed by myosin II motors. The simplest model for prestress in a
mechanical network is the introduction of a finite resting length which
is smaller than the typical extension of each link. Indeed, one- and
two-dimensional spring networks with prestress are widely used for
modeling cell migration~\cite{DiMillaLauffenburger91,Othmer,Bottino01,
Gracheva10}.

Network models are conceptually very appealing due to their
multi-scale nature: by changing the microscopic rules for the
mechanics of the links, one can explore how the macroscopic behavior
of the whole network changes. In particular, important biological
effects like viscoelasticity of the links or coupling to diffusion
fields can be incorporated~\cite{sureshbpj,sureshpnas,fedosov,
DiMillaLauffenburger91,Othmer,Bottino01,Gracheva10}. Spring networks offer
the additional advantage that homogenization techniques can be used
to arrive at continuum models. Recently, the interplay between force
generation and the geometrical and adhesive properties of the
environment have been addressed using the powerful framework of finite
element models (FEM) \cite{EvansPNAS,chen3d}, which can be
considered as the continuum limit of appropriate network models.
Most FEM-models use constitutive equations which correspond
to Hookean networks.

Although conceptually very appealing, modeling cell mechanics with
Hookean networks does not reflect the fact that the actin cytoskeleton
does not provide much resistance to compression. This is especially
true for two-dimensional networks for cell adhesion and migration,
because in this case the network might contract laterally, while the
cytosol flows into the third dimension. In this situation, the network
links do not behave as springs, but rather as cables, which are
characterized by an asymmetric force-extension relation. There are
several microscopic reasons for this effective behavior: not only do
thin actin bundles easily buckle under load, they also tend to
telescope in due to filament sliding and even to depolymerize once
tensile stress is relieved. Cable networks have been successfully used
to model the prestress-dependent mechanical response of adherent cells
to local mechanical perturbations~\cite{stamenovic}. The same model
has also been used to describe how mechanical stress is propagated
from the nuclear region through the cytoskeleton towards focal
adhesions, where changes in load lead to changes in adhesion
size~\cite{paul}. 

One striking feature of strongly adhering cells is the fact that
retracted contours often take the shape of circular arcs
\cite{c:zand89,c:barz99,c:lehn03,bischofsbpj08}.  Although for cable
networks the resulting shapes are strongly invaginated, it has been
shown that the circular invaginations observed for cells pinned at
discrete sites of adhesions can only be explained if an additional
contractile force is introduced for each mechanical
link~\cite{bischofsbpj08}. This additional force in an actively
contracting cable network does not vanish at the resting length and
represents the fact that contractility arises mainly from myosin II
motors, which in steady state operate close to a non-vanishing stall
force.

In this paper, we systematically compare the different network models
introduced before for adherent cells (spring models, cable networks,
and actively contracting networks) in regard to the predicted shapes
and force distributions. Our main conclusion is that actively
contracting cable networks share many crucial features with
adherent cells. Due to their linear nature, actively contracting
spring networks are equivalent to passive spring networks with a
reduced resting length.  Passive networks (both from springs and
cables) have a well-defined reference state even in the absence of
adhesion constraints and in general give similar results regarding
shape and force distribution, which is determined mainly by global
inputs like the spatial distribution of the adhesion points. In
contrast, actively contracting cable networks do not have a
well-defined reference state because without adhesion constraints,
they contract onto a point. In this case, we find that shape and force
distributions are determined mainly by the local distribution of
adhesion sites. The internal force distribution is constant in the
bulk and strongly localizes to the contour, where forces jump by
orders of magnitude. This motivates a detailed study of two contour
models, which allow us to derive analytical predictions which we
then compare with the results from the computer simulations.
We also discuss how actively contracting cable networks
can be extended to model non-linear or adaptive linker mechanics,
and comment on the relation of our network models to tissue mechanics.

\section{Network Models}

\subsection{Link Mechanics}

\begin{figure}
\begin{center}
\includegraphics[width=0.8\columnwidth]{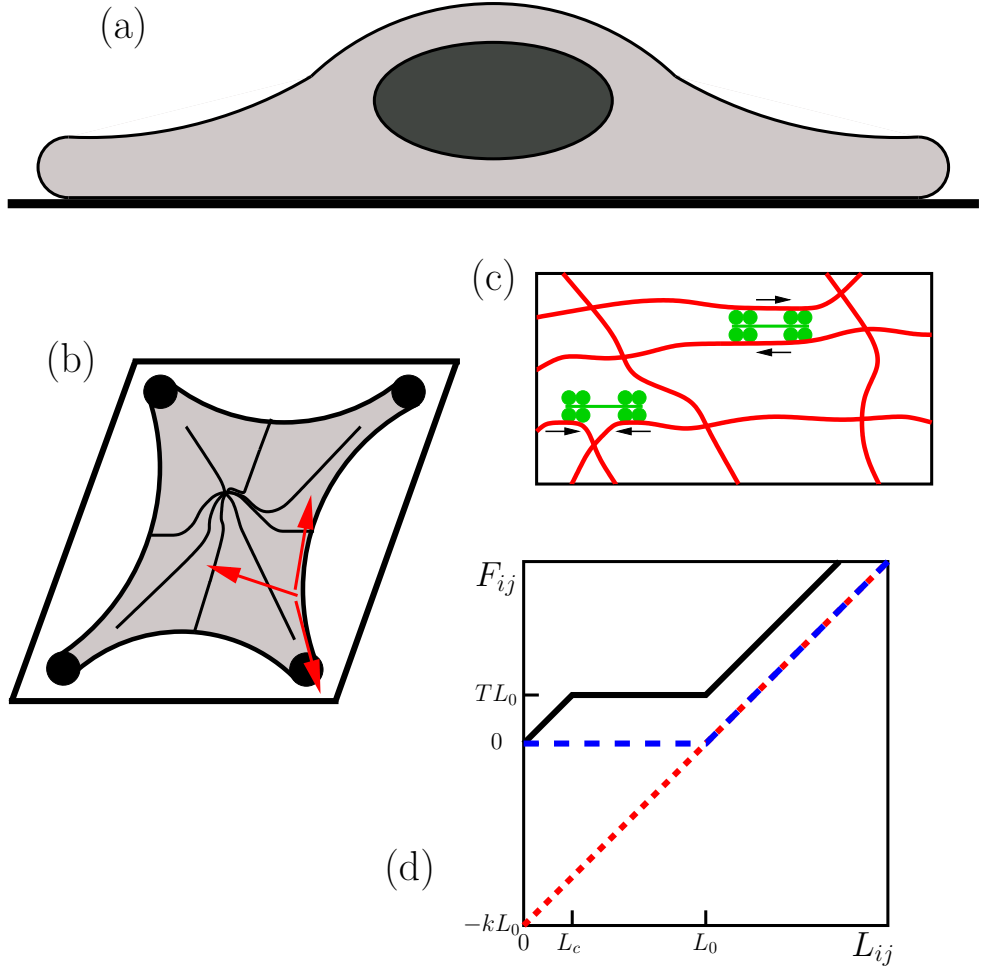}
\caption{(Color online) Sketch of the system. (a) Side-view of an
 adherent tissue cell.  (b) Top-view. The cell is assumed to be 
 adherent at four discrete dots. Its contour shows an invaginated 
 shape resulting from a balance between bulk forces
 directed towards the cell center and boundary forces directed
 along the cell periphery. (c) The actin cytoskeleton is tensed by 
 myosin II minifilaments, which actively contract the network with 
 forces $\vec{F}_{active}$.  If the network links are strained, 
 restoring forces $\vec{F}_{mech}$ appear. (d) Force-extension curve 
 $F_{ij}(L_{ij})$ of a link $ij$ in a simple Hookean network, a
 passive cable network, and an active cable network (in order of increasing dash
 lengths).  ($k$ = spring constant,
 $T$ = motor stall force per length, $L_0$ = initial link length,
 $L_c$ = critical length)}
\label{model}
\end{center}
\end{figure}

Tissue cells adhering to discrete adhesion points on a flat substrate
usually become very flat and therefore effectively two-dimensional.
Only the nucleus, which rises in the middle, makes them fully
three-dimensional, see sketches in \fig{model}a,b. However, here we
focus on lateral contraction and contour effects and thus the nucleus
is expected to play a minor role. In the following we therefore
restrict ourselves to two dimensions and model the cytoskeletal
network as a two-dimensional mesh of mechanical links joined at $N$
discrete nodes. Nodes are labeled with indices, e.g.\ $i$ and $j$. 
Microscopically links and nodes may represent filament bundles and
local accumulation of cross-linkers, respectively, but in a more
general sense, these mechanical elements are simple representatives of
an unknown network architecture which we model in a statistical sense.
The network is subject to internal forces originating from molecular
motor activity, $\vec F_{active}$, and the mechanical resistance of
filaments to strain, $\vec F_{mech}$, compare \fig{model}c. 

We introduce three fundamentally different kinds of mechanical models
for the network links. The simplest case is a Hookean spring network
(HSN) composed of links with resting length $L_0$, which represent
linear springs with spring constant $EA/L_0$, where $E$ is
the Young's modulus of the link and $A$ its cross-section.  The
restoring force acting on a node $i$ due to elastic strain in the link
$ij$ then reads:
\begin{equation}
\vec F_{ij,mech}= EA u_{ij}\vec e_{ij},
\end{equation}
where $\vec e_{ij}=(\vec R_{j}-\vec R_{i})/L_{ij}$ is the dimensionless unit
vector along the link $ij$. Here, $\vec R_{i}$ and $\vec R_{j}$ specify the node positions,
$L_{ij}$ is the length of the link and $u_{ij}=(L_{ij}-L_0)/L_0$ is
the strain in the link. The linear force-extension curve of a single link in
a HSN is shown in \fig{model}d as a dashed line with short dashes.

The mechanical properties of the cytoskeleton are attributed mainly to
the actin part. Actin is a semi-flexible filament prone to buckling
under compression and thus behaves like a cable, which can be
stretched but not compressed.  The Hookean assumption of a symmetric
elastic response is therefore not valid. The mechanical properties of
actin networks on a coarse-grained scale are more accurately described
by assuming a finite resistance of filaments to tension,
but no resistance to compression. The
mechanical restoring forces originating from a link connecting two
nodes $i$ and $j$ in the passive cable network (PCN) are therefore
given by:
\begin{eqnarray}
\label{pass_cables1}
\vec F_{ij,mech} &=& E A u_{ij} \vec e_{ij}, \quad{L_0<L_{ij}} \\
\vec F_{ij,mech} &=& 0, \quad{L_{ij}\leq L_0}.
\label{pass_cables2}
\end{eqnarray}
We show the asymmetric force-extension relation of the PCN links in \fig{model}d
as a dashed line with long dashes.

Let us assume that molecular motors are homogeneously distributed in
the network. Because they are arranged in a parallel fashion, 
their individual forces add up. We therefore assume that a link
contracts with a force $TL_0$ proportional to its initial length, where $T$
is force density per length. For computational simplicity, here we assume
that this force does not change as the filament contracts,
although in practise, it might well be that the line density
rather than the total number of active motors is constant. 
For a link $ij$ we therefore have:
\begin{equation}
\vec F_{ij,active}\text{ }=\text{ }TL_0 \vec e_{ij}\ ,
\end{equation}
where $T>0$ is the tensile force per initial length applied by the
motors. The finite force at zero length is unphysical and we avoid
it by introducing an additional rule such that force is diminished
if two neighboring nodes come closer to each other than some small
distance $L_c\ll L_0$:
\begin{equation}
\vec F_{ij,active}\text{ }=\text{ }TL_0 \frac{L_{ij}}{L_c} \vec e_{ij},
\qquad{L_{ij}<L_c}.
\end{equation}

Combining PCN and active contraction, we obtain what we call the active
cable network (ACN), compare the solid line in \fig{model}d:
\begin{eqnarray}
\vec F_{ij} &=& \left(TL_0 + E A u_{ij}\right)
  \vec e_{ij},\quad{L_0<L_{ij}},
  \label{act_cables1}\\
\vec F_{ij} &=& TL_0 \vec e_{ij},\quad{L_c\leq L_{ij}\leq L_0},
  \label{act_cables2}\\
\vec F_{ij} &=& TL_0 \frac{L_{ij}}{L_c} \vec e_{ij},\quad{L_{ij}<L_c}.
  \label{act_cables3}
\end{eqnarray}
Neither the detailed choice of $L_c$ nor the assumption of a linear
force reduction below $L_c$ are crucial for our results. 

Active contraction can also be combined with the HSN. However, this simply
shifts the straight dashed line in \fig{model}d,
i.e.\ this reduces the resting length $L_0$ to
\begin{equation}
L_0'=L_0\left(1-\frac{TL_0}{EA}\right) \label{red_leng}
\end{equation}
as  long as $TL_0\leq EA$. Therefore active contraction does not
change the basic definition of the HSN and therefore 
in the following the actively contracting HSN is not discussed further.

\subsection{Mechanical Equilibrium}

For nodes within the network, the total force exerted on a node $i$
is the sum of all forces applied by neighboring nodes $j$:
\begin{equation}
\vec F_{i} = \sum_{j} \vec F_{ij}.
\end{equation}
In mechanical equilibrium, the force on each non-adherent node has to
vanish, $\vec F_{i}=0$. The existence
of adhesion sites is modeled by fixing the positions of the adherent
nodes.  Thus the adhesion site geometry will enter through the
boundary conditions.

To reduce the number of parameters, we scale all lengths with respect
to $L_0$, e.g.\ $\ell=L/L_0$. All forces are scaled as $f=F/EA$. We
define the ratio of active to elastic forces as
\begin{equation}
\tau = \frac{TL_0}{EA}\ .
\label{act_elast}
\end{equation}
For an ACN we therefore may rewrite the forces acting on a node in the
non-dimensionalized form as:
\begin{eqnarray}
\vec f_{ij}&=& (u_{ij}+\tau) \vec e_{ij}
  \qquad{1<\ell_{ij}}, \\
\vec f_{ij} &=& \tau \vec e_{ij}
  \qquad{\ell_c\leq \ell_{ij}\leq1},\\ 
\vec f_{ij} &=& \tau \frac{\ell_{ij}}{\ell_c} \vec e_{ij}
  \qquad{\ell_{ij} < \ell_c}.
\end{eqnarray}
In the computer simulations we use $\ell_c=10^{-3}$.

Mechanical equilibrium requires the forces on each non-adherent node
to vanish
\begin{equation}
\sum\limits_{j}\vec{f_{ij}} = 0 
  \qquad\forall{\text{ non-adherent nodes }i}
\label{system}
\end{equation}
with the summation $j$ over all neighbors.
For a two-dimensional network of $N'$ non-adherent nodes, the system of equations
(\ref{system}) consists of $2N'$ coupled non-linear equations. If the
lhs of system (\ref{system}) consisted of arbitrary functions of the
$\vec r_{i}$, the method of choice to solve it would be the
Newton-Raphson method ~\cite{geraldwheatley}.  However, since the lhs
of system (\ref{system}) is a force which has a potential, it is also
a $2N'$-dimensional gradient vector.  Therefore we solve the
minimization problem for the potential with the conjugated gradient
method~\cite{fletcherreeves}. We stop iterating as soon as the force
on every node (except the periphery nodes) is smaller by at least two
orders of magnitude than the smallest link force.

\subsection{Parametrization}

Due to its dynamic and multiscale organization and the known
limitations of microscopy, a detailed model of the actin cytoskeleton
is currently out of bounds. In the face of these uncertainties, our
model is not meant to represent the details of the organization of the
actin cytoskeleton.  Nevertheless, for practical purposes it is
helpful to parametrize our model using some benchmark values for the
actin cytoskeleton.

The elastic modulus of an actin filament, which has cross-section area
$A_{fil}=18.8$ $nm^2$, was experimentally found to be $E_{fil}=2.8$
$GPa$~\cite{mickey}, while typical values for stress fibers are
a radius around $100$ $nm$ (corresponding to $A_{fib}=31416$ $nm^2$)
and an effective modulus of $E_{fib}=1.45$ $MPa$~\cite{deguchi}. Hence,
the Young's modulus of stress fibers is three orders of magnitude
smaller than the one of single actin filaments. This suggests that
cross-linkers like $\alpha$-actinin and myosin II are the main
contributors to elasticity and not the actin filaments
themselves. However, the values for the one-dimensional 
modulus, $E_{fil}A_{fil}=52.6$ $nN$ and
$E_{fib}A_{fib}=45.6$ $nN$, are effectively very similar, so the
one-dimensional modulus is expected to be of the order of $50$
$nN$.

The mesh size of the cytoskeleton is expected to be typically around
$L_0=100$ $nm$. This is an intermediate value introduced
in~\cite{stamenovic} based on experimental observations of the actin
cytoskeleton in adherent endothelial~\cite{satcherdewey} and
fibroblast cells~\cite{lubyphelps}. For the active force, we estimate
that around 1.000 myosin II motors are active in one effective
link. With a stall force of 2 pN per motor head \cite{molloy}, we have
$T=2\cdot10^{-2}\ nN/nm$ for the motor force per length. Using
\eq{act_elast} and $EA=50$ $nN$, we estimate $\tau = 0.04$ for the
active tension in the network.

\section{Results}

\subsection{Equilibrium Shapes}

\begin{figure}
\begin{center}
\includegraphics[width=0.8\columnwidth]{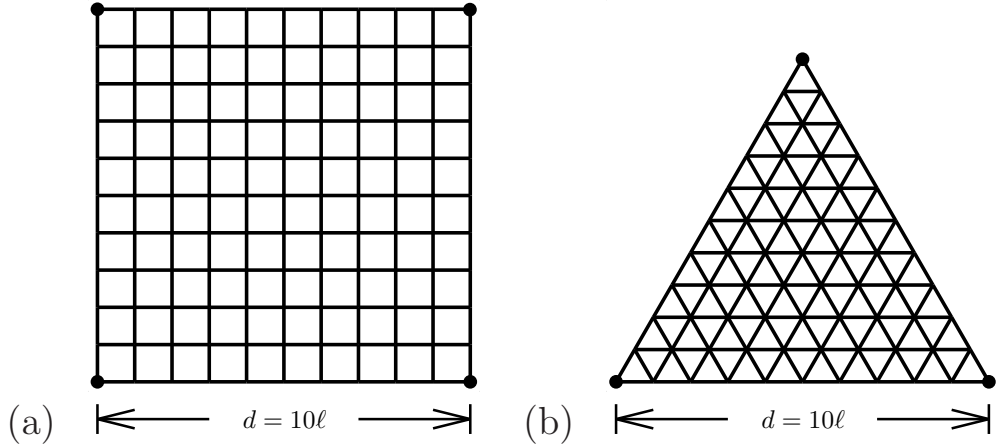}
\caption{Tension-free reference states. (a) Square network with link
 length $\ell=1$ and side length $d=10\ell$. (b) Triangular network
 with the same $\ell$ and $d$.}
\label{ref_states}
\end{center}
\end{figure}

\begin{figure}
\begin{center}
\includegraphics[width=0.8\columnwidth]{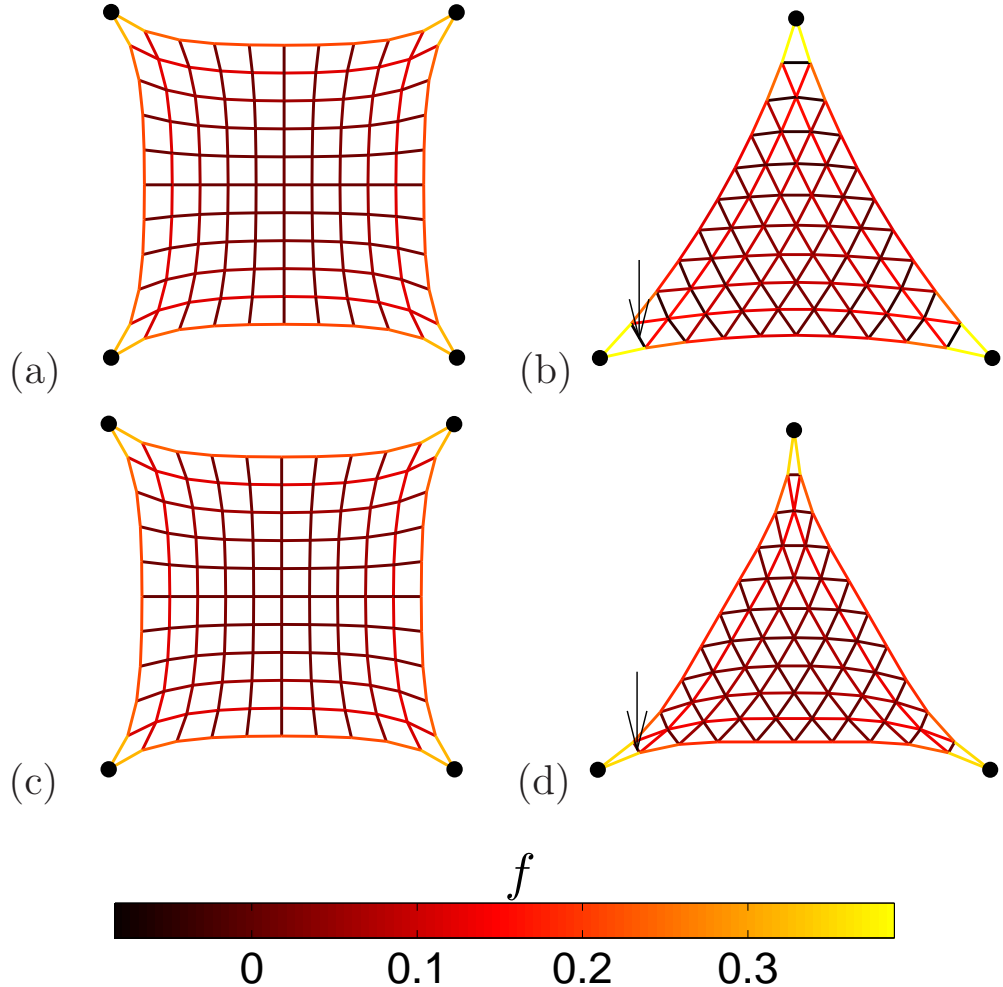}
\caption{(Color online) Tensed Hookean (HSN) and passive cable networks (PCN) with
 $\tau_H=0.2$. The colorbar gives the dimensionless force $f$. (a)
 Contraction of a HSN in the square reference state. (b)
 Equilibrium shape of a HSN in triangular reference state. (c) PCN in
 square geometry. (d) PCN in triangular geometry. For the triangular
 reference shape, HSN (b) and PCN (d) differ in the amount of
 compression in the thin extensions (marked by arrows).}
\label{hook_pass_contr}
\end{center}
\end{figure}

In the following we analyze how the different types of networks act
under tension for given adhesion constraints. We first discuss
passive networks. In order to understand
the role of geometry, as tension-free reference states we use both a
square and a triangle as shown in \fig{ref_states}. If the corners are
fixed and the network is set under tension by reducing rest length, invaginated shapes appear
as shown in \fig{hook_pass_contr} for Hookean spring networks (HSN)
and passive cable networks (PCN). Contraction is
quantified by $\tau_H=(\ell-\ell_0)/\ell$. In
\fig{hook_pass_contr}, the results for HSN and PCN are the same for
the square shape, \fig{hook_pass_contr}a,c, because these two kinds of
networks behave identical as long as all links are tensed.  However,
the results are different for the triangle shape,
\fig{hook_pass_contr}b,d. In this case, the PCN gives a significantly
flatter contour due to the missing response to compression in the thin
extensions leading to the adhesion points.  The most prominent examples for
compressed links in \fig{hook_pass_contr}b,d are indicated by arrows.

In order to understand our numerical results in more detail, we first
note that the HSN with triangular network topology has a well-defined
continuum limit, in which it corresponds to a two-dimensional sheet
with isotropic linear elastictiy \cite{boal,hansen1}. The two
corresponding elastic constant are a Young's modulus of $2 k /
\sqrt{3}$ and a Poisson's ratio of $1/3$, where $k = EA/L_0$ is the
spring constant of the links. The HSN with simple cubic topology does
not have such a rigorous limit, but in our context it works in a
similar way as the triangular lattice. Therefore similar results as
obtained here for the HSN are also obtained with continuum elasticity
theory applied to two-dimensional cell shapes
\cite{EvansPNAS}. Without any adhesion constraint, the HSN contracts
isotropically to a finite size, i.e.\ the network is uniformly scaled
and has a new side length $d'=10\ell_0$. This shape we call the
unconstrained reference shape and it is key to understand the
results for HSN. The same shape as shown in \fig{hook_pass_contr}a
results if the network contracts away from its initial state under
adhesion constraints or if the network starts from its unconstrained
reference state and its corners are dragged to the desired adhesion
points. This explains the main feature of the force distribution
shown by the color coding in \fig{hook_pass_contr}a, namely the strong
localization of stresses and strains to the regions around the adhesion
points. With the amount of tension used here, the network can attain
its unconstrained reference state away from the adhesion points and
therefore its contour is essentially flat in the middle parts.

\begin{figure}
\begin{center}
\includegraphics[width=0.8\columnwidth]{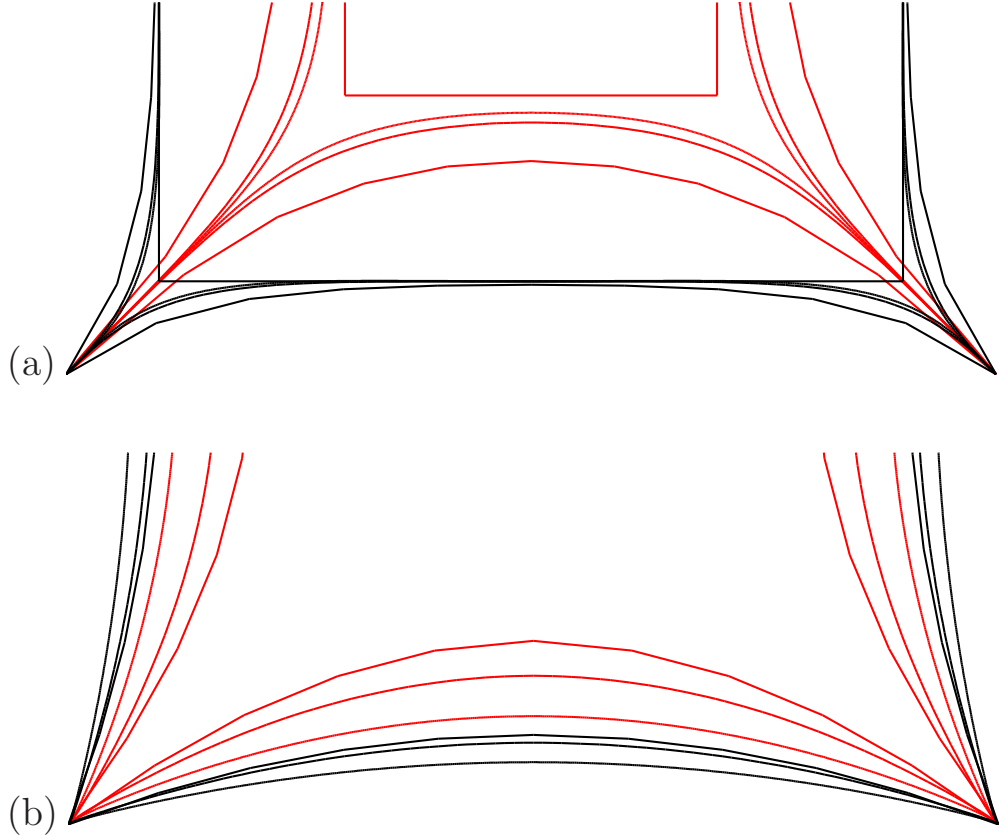}
\caption{(Color online) (a) Boundary line of the HSN or PCN for the square shape
 for network tension $\tau_H=0.2$ (lower set) and $\tau_H=0.6$ (upper set).
 The straight line shows the unconstrained reference shape, while the three
 curved lines represent different initial link lengths, namely
 $\ell=1,\text{ }0.1,\text{ }0.02$ (from bottom to top within one
 set). (b) Boundary line of the active cable network (ACN) for the square shape
 for $\tau=10^{-3}$ (lower set) and $\tau=10^{-2}$ (upper set).
 In this case, no unconstrained reference shape exists. The
 three lines again represent the initial link lengths $\ell=1,\text{ }
 0.1,\text{ }0.02$ (from top to bottom).}
\label{contour_hook_cable}
\end{center}
\end{figure}

In \fig{contour_hook_cable}a we directly compare the calculated network
shapes for the HSN for the square geometry to the unconstrained reference
shape. In addition, we demonstrate the role of the link length $\ell$.
As explained above, in this case the PCN gives the same 
results. For the small value of tension, $\tau_H = 0.2$ (lower set), the
contracting network can reach the unconstrained reference shape over a
large region where it is therefore essentially flat. The smaller $\ell$, the
faster this contour is reached due to an increased force density along
the contour. For the large value of tension, $\tau_H = 0.6$ (upper set), the
unconstrained reference shape is not reached by the contracted network
and it stays non-flat along its whole contour even for rather small
values of $\ell$.

\begin{figure}
\begin{center}
\includegraphics[width=0.7\columnwidth]{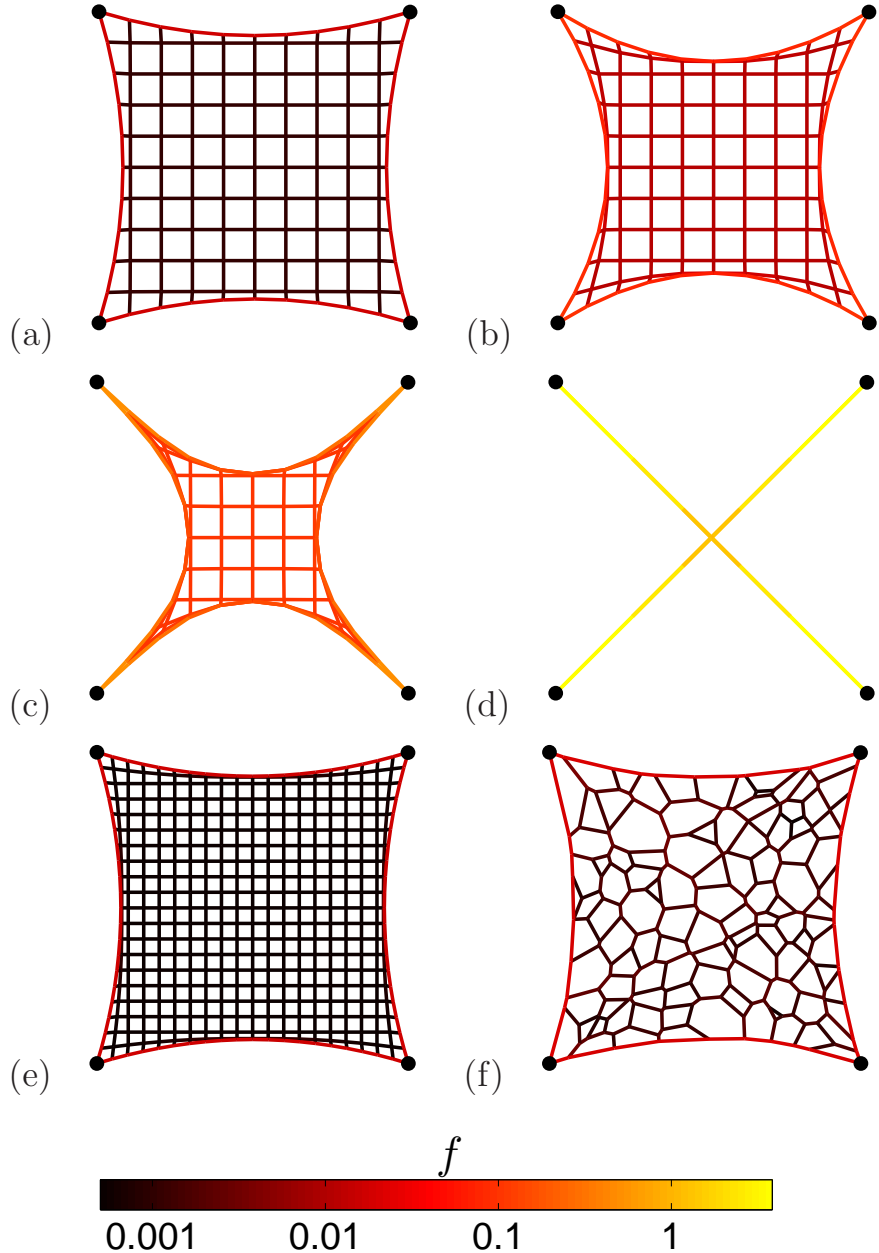}
\caption{(Color online) Contraction of an active cable network (ACN) with square shape.
  (a) and (b) Contraction of the network leads to arc formation.
  (c) With increasing tension tubes form near the adhesion points.
  Note the bundling of filaments at the edge.
  (d) For very large $\tau$, the network collapses onto the center tree.
  (e) The same situation as in (a) but with $\ell=0.5$. The contour is equal
  to that in (a) and the boundary forces are the same.
  (f) Square Voronoi network with 212 nodes and 316 links.
  Note the regular contour and the strong stress localization in the periphery.
  Tension values are (from (a) to (f)):
  $\tau=10^{-3},\text{ }10^{-2},\text{ }10^{-1},\text{ }1,\text{ }10^{-3},
  \text{ }10^{-3}$.}
\label{cable_contr_hom}
\end{center}
\end{figure}

Next we turn to actively contracting networks. 
Figs.\ \ref{cable_contr_hom}a-d show the equilibrium shapes of an
active cable network (ACN) with reference state from
\fig{ref_states}a and increasing tension $\tau$. As tension increases,
the shape becomes more and more invaginated, until it collapses onto
the zero-area network in \fig{cable_contr_hom}d. This network
basically consists of a centrally contracted region which is connected
to the adhesion points by long extensions. We therefore call this network
the center tree (CT). Note that this network still retains aspects of
the two-dimensional network, because an effectively one-dimensional
structure would collapse onto the so-called Steiner tree of minimal
length, which for a square shape is not four-fold symmetric \cite{parsimonious}.

In contrast to the passive networks, where tension ceases as the
unconstrained reference state is reached, for the ACN no such
unconstrained reference shape exists and without adhesion constraints
the shape would collapse onto a single point.  This explains why
flat parts are not observed in the contours of the networks
shown in \fig{cable_contr_hom}a-d. This is also demonstrated
in \fig{contour_hook_cable}b, where we show contour shapes 
for $\tau=10^{-3}$ (lower set) and $\tau=10^{-2}$ (upper set).
Like for the HSN in \fig{contour_hook_cable}a, increasing
lattice constant $\ell$ increases the force density along the contour
and therefore leads to stronger invagination. However, in marked
contrast to the HSN, no flat parts appear in the contour as
no unconstrained reference state exists.

\fig{cable_contr_hom}a-d also shows that the formation of inward directed
arcs now corresponds to a much more inhomogeneous density distribution
of filaments: in the bulk of the network the distance between nodes
and thus the filament density remains unchanged, while at the edges
filaments start to bundle strongly along the edge. The color code in
\fig{cable_contr_hom} shows that stress is strongly localized at the
periphery. In the interior, the only forces acting are
the motor forces $\tau$ which balance each other at every node. 
At the periphery, the force jumps up from $\tau$ to much higher values
$\tau+EA\Delta\ell_j/\ell$, compare
\fig{cable_contr_hom}a. The forces are largest close to the
adhesion points and decrease towards the center of the boundary. 
For large tension $(\tau > 10^{-2})$, tubes are formed near the
adhesion points and the stress distribution along the contour
becomes more inhomogeneous. 

\fig{cable_contr_hom}e shows the effect of changed discretization for
the same tension value as \fig{cable_contr_hom}a for 
the complete network. \fig{cable_contr_hom}f demonstrates that for
ACN, shape and force values do not depend significantly on network
topology. As an instructive example here we use a disordered network
topology obtained by a Voronoi construction. Even the presence of
relatively large elements in the discretization does not change the
invaginated shape feature of the contracted network. We conclude from
\fig{cable_contr_hom}e and \fig{cable_contr_hom}f that ACN are
surprisingly robust in regard to the details of the network topology.
In general, similar results as obtained in \fig{cable_contr_hom} for the square
shape are also obtained for other initial cell shapes.

\begin{figure}
\begin{center}
\includegraphics[width=0.8\columnwidth]{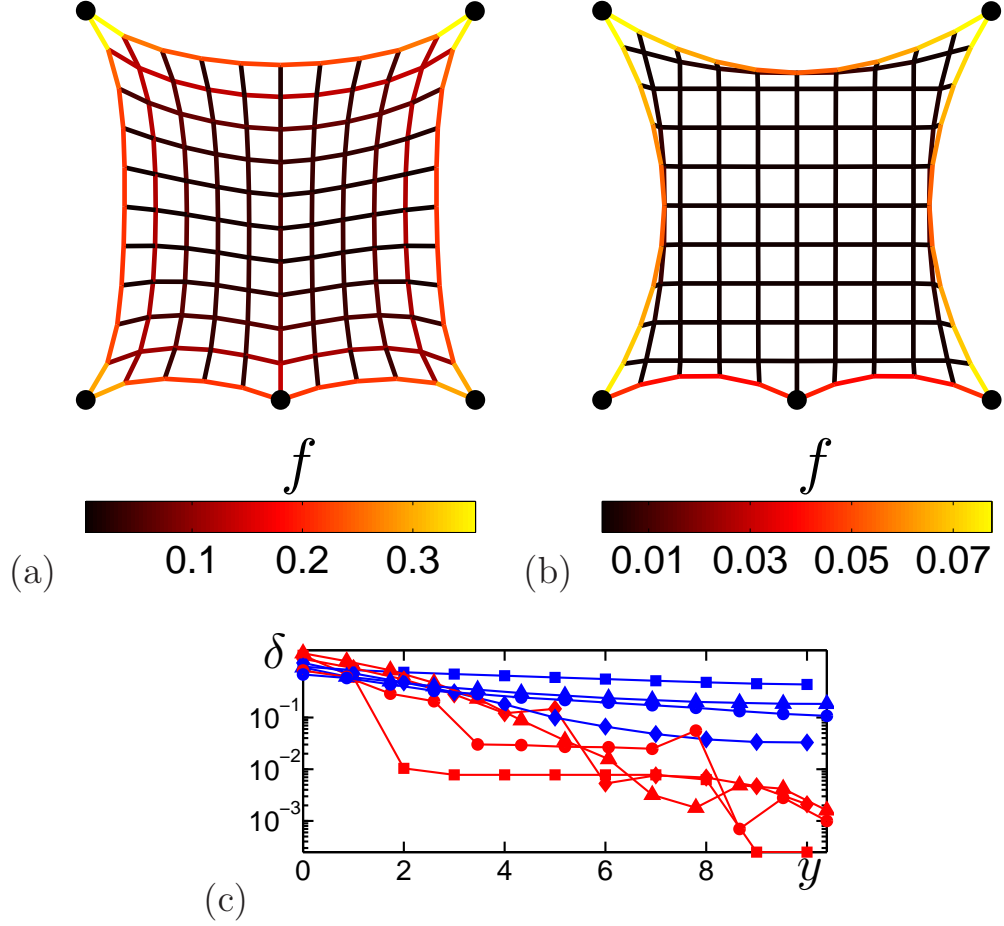}
\caption{(Color online) Contraction of networks with square shape
 with an additional adhesion point in the middle of the bottom line.
 (a) HSN or PCN model with $\tau_H=0.2$. (b) ACN with $\tau=10^{-2}$. (c)
 Relative displacement $\delta$ of nodes with $x=5$ (vertical middle line) 
with and without the additional adhesion point at the bottom. $y$ is the
 node's $y$-coordinate in the initial network. The four top
 lines correspond to the HSN while the four bottom lines represent the ACN.
 Different symbols show different topologies (square=$\square$,
 diamond=$\Diamond$, triangular=$\bigtriangleup$, hexagonal=$\bigcirc$.)}
\label{coupling}
\end{center}
\end{figure} 

In \fig{coupling}, we investigate another striking property of ACN, namely its
robustness in regard to addition of new adhesion points.  In
\fig{coupling}a,b we show the equilibrium shapes of the HSN from
\fig{hook_pass_contr}a and the ACN from \fig{cable_contr_hom}b with
one adhesion point added in the middle of the bottom line.
In the case of the ACN model, this
change at the bottom of the network has little influence on the
positions of nodes not directly connected to the bottom line. In
contrast, for the HSN the additional adhesion point affects the shape
of the opposite arc, becoming more curved in the center. In
\fig{coupling}c we plot by which distance $\delta$ the nodes in the
vertical middle line are pulled down in the negative y-direction upon
addition of the new adhesion point. The plot of $\log \delta$ versus
$y$ is not smooth for numerical reasons, but clearly shows that the
effect decays much more rapidly for the ACN versus the HSN. In
addition to square and triangular network topologies, here we also
show results for rotated square (diamond) and hexagonal networks.
Intriguingly, stress in the contour behaves very differently, compare
the color coding of \fig{coupling}a,b.  While in the HSN stress in the
bottom line stays approximately the same, in the ACN it decreases to
half its value, indicating a strong effect on contour forces.

In summary, ACNs behave very differently from HSNs (and therefore also
from the mostly equivalent PCNs). Roughly speaking, they respond more
locally than globally. They are more robust in regard to network
topology and adhesion geometry and show strong localization of the
stress to the periphery. Addition of new adhesion points leads to
little global change, but to a strong change in local stress distribution.
Because for ACN shape and contour stress seem to be mainly
determined by the local adhesion geometry, they will now be analyzed in
more detail. 

\subsection{Contour shape and tension-elasticity model}

\begin{figure}
\begin{center}
\includegraphics[width=0.8\columnwidth]{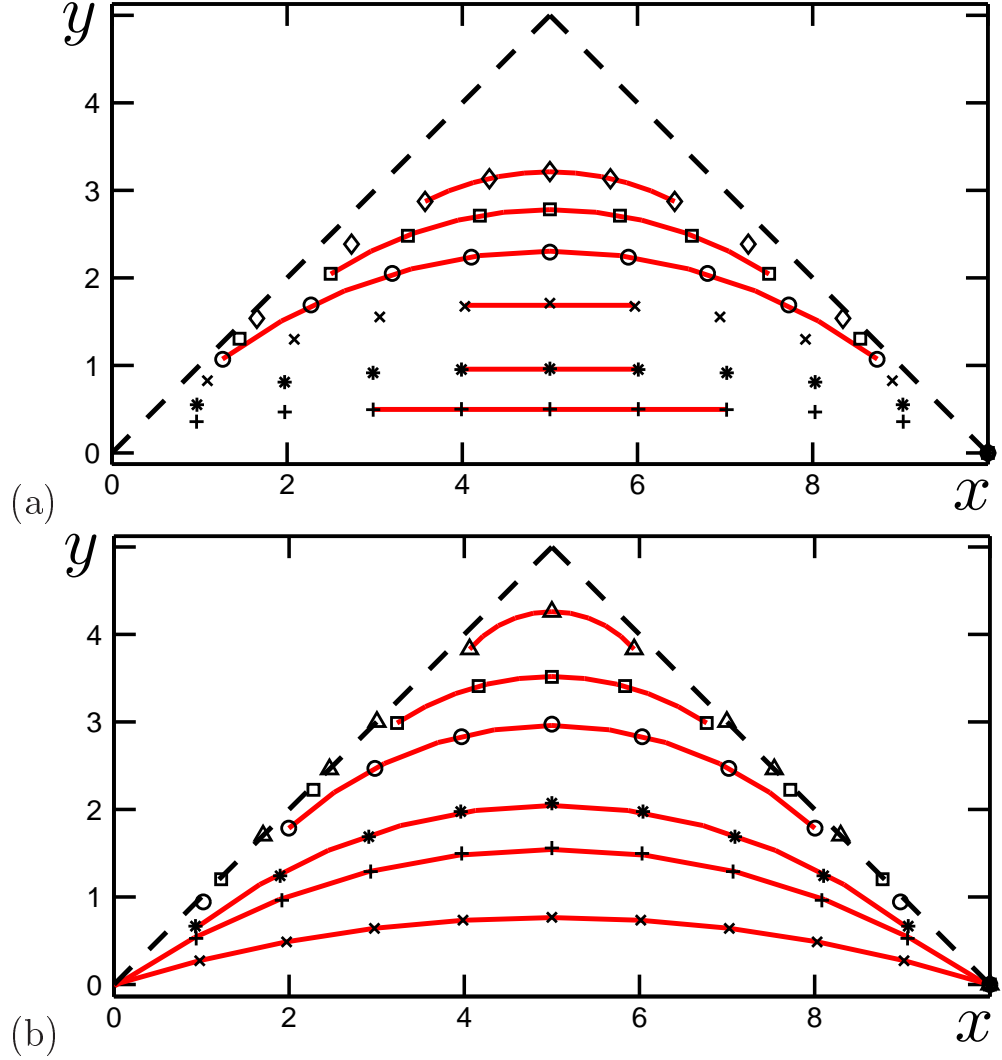}
\caption{(Color online) Arc fits. (a) Arc analysis for
  a tensed HSN or PCN. From bottom to top: $\tau_H=0.1$ ($+$), $0.2$ ($*$),
  $0.4$ ($\times$), $0.6$ ($\bigcirc$), $0.8$ ($\square$), $1$
  ($\Diamond$). Symbols: Contour. Lines: Straight fits of the linear
  contour parts (the three bottom lines), least square fits of round
  contour parts to arcs with constant curvature (the three top lines).
  $*$ corresponds to the bottom line from the networks shown in
  Figs.\ \ref{hook_pass_contr}a and c. (b) Contour analysis for the
  bottom line of an ACN with $\tau=10^{-3}$ ($\times$), $10^{-2}$
  ($+$), $2.5\cdot10^{-2}$ ($*$), $10^{-1}$ ($\bigcirc$),
  $2.5\cdot10^{-1}$ ($\square$), $5\cdot10^{-1}$ ($\bigtriangleup$)
  from bottom to top.  Symbols denote node positions of the network
  arcs while the lines are least square fits of the contours to arcs
  of constant curvature. $\times$, $+$, $\bigcirc$, and dashed line
  correspond to the bottom lines from
  Figs.\ \ref{cable_contr_hom}a-d.}
\label{cable_arcs}
\end{center}
\end{figure}

In contrast to HSNs, the contour of ACNs appears to be more
circular. Indeed, a circular arc morphology has been noted before for
the shapes of cells adhering to micropatterned substrates and
therefore this shape feature is an important motivation to study
ACNs~\cite{bischofsbpj08}. We now investigate this important aspect in
more detail. In \fig{cable_arcs}a and b we show contours of the HSN
and ACN from \fig{hook_pass_contr} and \fig{cable_contr_hom},
respectively, together with fits to straight lines or circular
arcs. As we increase tension, we observe more invaginated shapes,
compare \fig{contour_hook_cable}. For the HSN in \fig{cable_arcs}a,
small tension allows the network to reach the
unconstrained reference shape and therefore the best fit to the middle
part of the contour is a straight line. For larger tension, the unconstrained
reference shape cannot be reached anymore and circular shapes become
better fits. This crossover is in marked contrast to the ACN from
\fig{cable_arcs}b, where circular arcs fit very well for all values of
$\tau$. For large $\tau$, the overall contour starts to deviate from
the perfect arc shape because the networks starts to collapse into
tubes near the adhesion points. However, locally (in between
the tubes) the contour stays circular.  Another difference between the
two network types lies in the observation that for HSN, network shape
strongly depends on lattice constant $\ell$, while for ACN, the
equilibrium contour is relatively independent of $\ell$.

\begin{figure}
\begin{center}
\includegraphics[width=0.8\columnwidth]{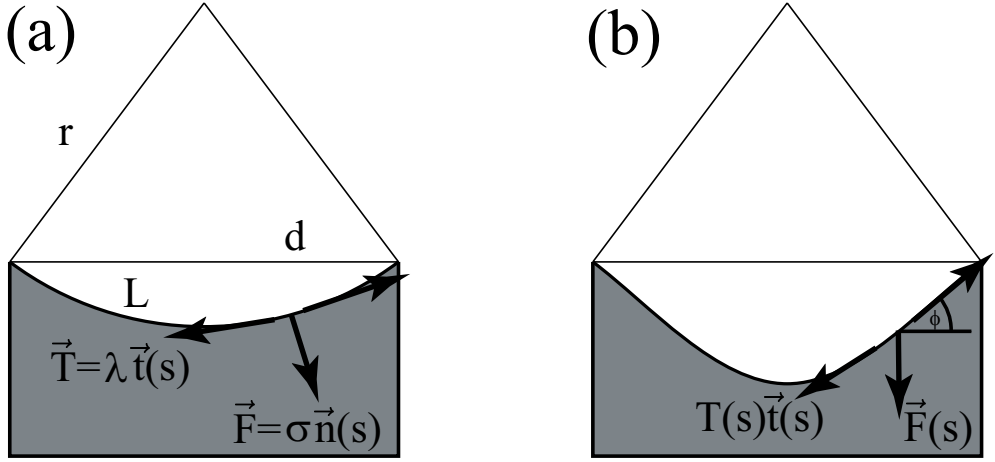}
\caption{Schematic representation of the two contour models. (a) In
  the tension-elasticity model (TEM), an isotropic surface tension
  $\sigma$ pulls the contour in along the normal direction, while the
  counteracting line tension $\lambda$ acts along the tangential
  direction. $r$ is arc radius, $L$ is contour length, and $d$ is
  spanning distance. (b) In the elastic catenary model, the inward
  pull is vertical and thus leads to an inhomogeneous line density of
  force along the elastic contour. $\phi$ denotes the tangential angle.}
\label{cartoon}
\end{center}
\end{figure}

It has been argued before that the circular arc shape feature of the
ACNs can be explained by an analytical theory, the tension-elasticity
model (TEM) \cite{bischofsbpj08}. For clarity, here we repeat this
analysis and compare it in detail with our network
simulations. Because ACNs do not propagate compression and the motor
forces represent a constant pull in the network, in the TEM the bulk
contractility is modeled by a structure-less surface tension
$\sigma$. However, elasticity is crucial to understand how the contour
reacts to the internal pull.  Therefore the elastic nature of the
mechanical network is represented by an elastic line tension
$\lambda$, which prevents the contour from collapsing under the inward
pull of the bulk network. This line tension is written as
\begin{equation}
\lambda=EA \frac{L-L_0}{L_0}
\label{ElasticLineTension}
\end{equation}
where $L$ is the contour length and $L_0$ is its resting length. Note
that we use the same value $EA$ like for the single links because the
elastic line tension will be dominated by the contribution from the
most peripheral line of links. We further assume $L_0 = \alpha d$,
where $d$ is the initial (\textit{spanning}) distance $d$ between two
neighboring adhesion points (which in the simulations above has been
chosen to be $10$) and $\alpha$ is a dimensionless resting length
parameter (compare Fig.~\ref{cartoon}a for a schematics). In the
following we restrict ourselves to $\alpha=1$, that is we assume that
a completely relaxed contour is straight, but without internal
tension. This implies that we neglect the contribution of the active
contractility in the periphery to the line tension.

The relation between surface tension $\sigma$ and network tension
$\tau$, which depends on network topology and discretization, can be
obtained numerically. For this purpose, we simulate the pulling of a
rectangular sheet of network. Surface tension $\sigma$ then follows as
total force on the pulling boundary divided by its width.  Due to this
normalization, the result does not depend much on link length $\ell$,
thus we use $\ell = 1$ for the simulations. For all
considered network geometries, we find a linear relation:
\begin{eqnarray}
\sigma_{square}&=&0.9907\cdot\tau_{square},\\
\sigma_{triangular}&=&1.6892\cdot\tau_{triangular},\\
\sigma_{diamond}&=&1.0867\cdot\tau_{diamond},\\
\sigma_{hexagonal}&=&0.5517\cdot\tau_{hexagonal}.
\label{sigma_tau2}
\end{eqnarray}
The constant for the square lattice is close to 
$1$ because here all links pull essentially perpendicular 
to the boundary.

\begin{figure}
\begin{center}
\includegraphics[width=0.8\columnwidth]{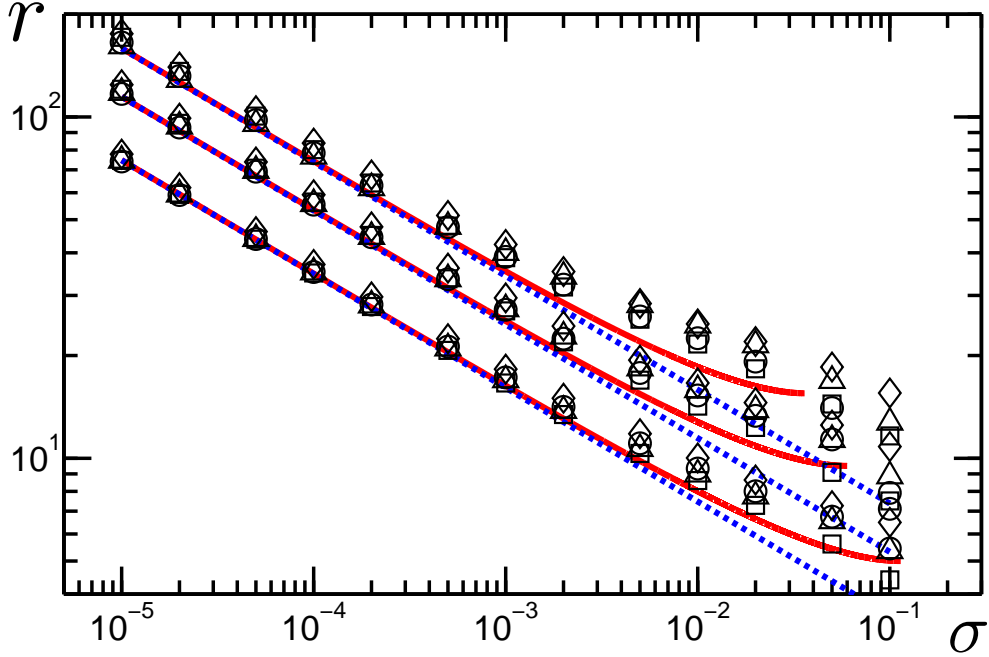}
\caption{(Color online) Relation of arc radii to internal tension of ACN for
 different dot distances and comparison to the tension-elasticity
 model (TEM). Symbols denote simulation results ($\square$ corresponds to
 square, $\Diamond$ to diamond, $\bigtriangleup$ to triangular, and
 $\bigcirc$\hspace{0.5mm} to hexagonal topology), the solid line is
 the numerical solution of \eq{tem} and the dashed line
 the analytical result \eq{tem_simple}. Side lengths are (from bottom to top):
 $d=10$, $19$, $31$, with critical tensions $\sigma_c\approx0.114$,
 $0.060$, $0.036$.}
\label{r_vs_d}
\end{center}
\end{figure} 

Given the forces assumed by the TEM, one can derive the shape of the
contour from the force balance.  While the surface tension $\sigma$
acts in the direction of the normal $\vec{n}$, resulting in 
a pulling force $\vec{F} = \sigma \vec{n}$, the line tension
$\lambda$ acts in the tangential direction $\vec{t}$ (compare
Fig.~\ref{cartoon}a). Because the elastic line tension is a global
quantity, it does not vary with the contour length $s$ and therefore
the contour tension is $\vec{T}(s) = \lambda \vec{t}(s)$. Then the
force balance reads
\begin{equation}
\vec{F} = \sigma\vec{n}= \frac{d\vec{T}}{ds} = \lambda\frac{d\vec{t}}{ds}=\frac{\lambda}{r}\vec{n}
\end{equation}
where for the second part we have used the geometrical relation
$d\vec{t}/ds = \vec{n} / r$ with $r$ being the radius of curvature. We
thus conclude that the TEM predicts circular arcs with a radius
\begin{equation}
r=\frac{\lambda}{\sigma}\ .
\label{LaplaceLaw}
\end{equation}
Although this results looks like a simple Laplace law in
two dimensions, it is more complicated, because the arc
radius $r$ will depend on global properties like spanning
distance $d$ through the elastic line tension $\lambda$
from \eq{ElasticLineTension}.

In order to arrive at an expression for arc radius $r$ as a function
of adhesion geometry and network tension, we use the trigonometric
relation 
\begin{equation}
\sin\left(\frac{L}{2 r}\right)=\frac{d}{2 r}
\end{equation}
to replace contour length $L$ by spanning distance $d$, compare
Fig.~\ref{cartoon}a. In combination with \eq{ElasticLineTension} (in dimensionless form) and
\eq{LaplaceLaw}, this gives
\begin{equation}
\label{tem}
r=\frac{1}{\sigma}\left(\frac{2r}{d}\arcsin
\left(\frac{d}{2r}\right)-1\right).
\end{equation}
Since this equation
cannot be solved analytically for $r$, it has to be solved numerically
for given values of $d$ and $\sigma$. For geometrical reasons, $r$ must always be larger than
$d/2$. Therefore, a critical $\sigma_c$ exists above which \eq{tem}
cannot be solved anymore. For small values of $\sigma$,
the invagination is small and one can expand the geometrical
relation in $d/r\ll1$. This leads to the analytical result
\begin{equation}
\label{tem_simple}
r=24^{-\frac{1}{3}}d^{\frac{2}{3}}\sigma^{-\frac{1}{3}}.
\end{equation}
In \fig{r_vs_d} we compare the results from computer simulations for
$r$ over a large range of network tension $\tau$ to the results of the
TEM with the corresponding range of surface tension $\sigma$, both for
the numerical solution of \eq{tem} and the analytical solution
\eq{tem_simple}. We note that the predicted power law behavior
applies over a very large range of tensions, and only breaks down at
very large tension $\sigma>10^{-2}$, compare \fig{r_vs_d}.
The inverse relation between $r$ and $\sigma$
represents a modified Laplace law for ACNs and thus demonstrates that
the concept of an isotropic surface tension works well to explain cell
shape. With the linear relation between $\sigma$ and $\tau$, this
implies that $r\sim\tau^{-\frac{1}{3}}$. 

We also find excellent agreement between computer
simulations and TEM upon variation of spanning distance,
compare \fig{r_vs_d}. Thus the elastic effects mediated by the spanning
distance $d$ are captured well by the concept of an elastic line
tension. In summary, the analytical TEM results in a surprisingly good
description of the contour shape of ACNs. As we will discuss
in the next section, however, agreement is less good regarding
contour forces.

\subsection{Contour forces and elastic catenary model}

\begin{figure}
\begin{center}
\includegraphics[width=0.8\columnwidth]{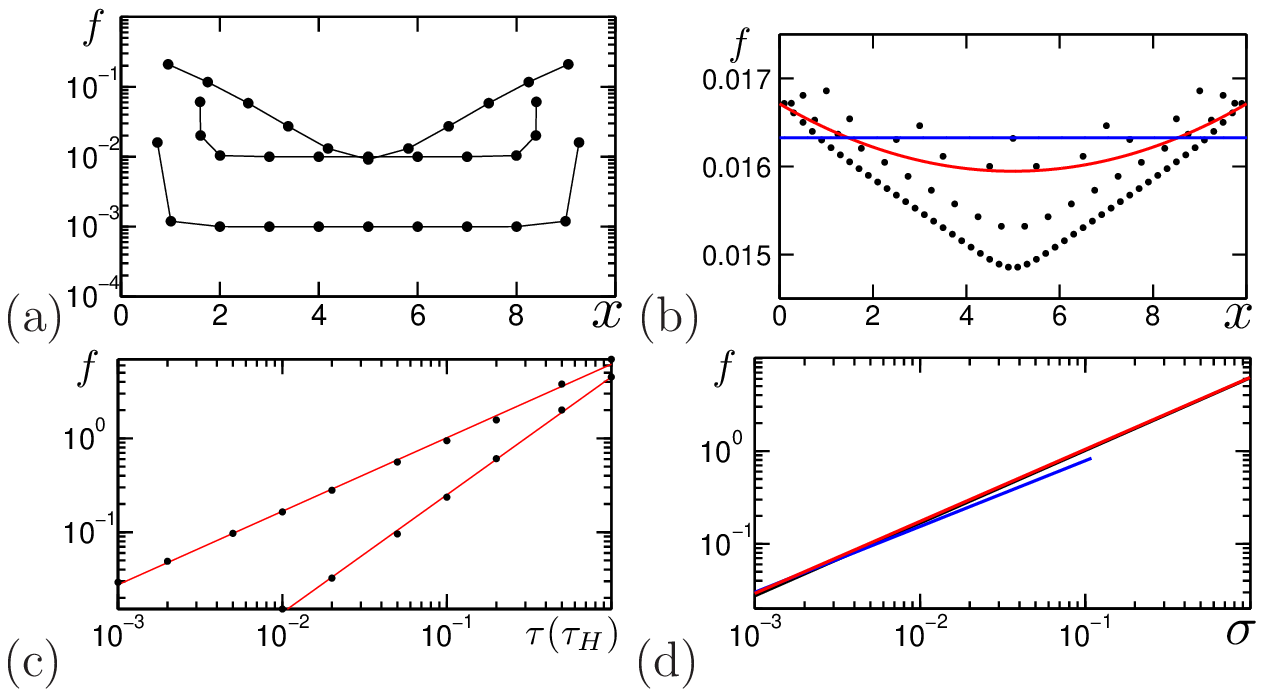}
\caption{(Color online) Force distribution in adherent networks. (a)
 Force of vertical links which cross the straight line $y=5.5$ in the
 HSN from \fig{hook_pass_contr}a (top) and the ACNs from
 \fig{cable_contr_hom}a,b (bottom, middle). (b) Force in the bottom
 line links of an ACN with $\sigma=10^{-3}$ for different lattice
 constants (symbols). From top to bottom: $\ell=2\text{, }1\text{, }
 0.5\text{, }0.2$. The straight line gives the line tension obtained
 via the TEM, the curved line follows from \eq{ten_c}. In both, (a)
 and (b), on the $x$-axis we have the $x$-coordinate of the center of mass
 of the links. (c) Forces on adhesion dots exerted by the ACN (top)
 and HSN (bottom). Simulation results are shown as dots, while
 lines give the power law fits. For the ACN we obtain $f=a\tau^b$
 with $a\approx6.16$ and $b\approx0.783$ and for the HSN
 $f=a\tau_H^b$ with $a\approx4.51$ and $b\approx1.26$. (d) Power law
 fits of adhesion dot force vs. surface tension. ACN and elastic catenary
 results (top line, collapsed) and TEM (low line).}
\label{stressplots}
\end{center}
\end{figure}

We now discuss the forces resulting from our computer simulations.
In the simulated networks of \fig{hook_pass_contr} for HSN and
\fig{cable_contr_hom} for ACN, the stresses in the network
are coded by colour. In \fig{stressplots}a we plot these stresses
along a line horizontally crossing
the networks with square shapes slightly above the middle line.
In the HSN, stress gradually decays into the sample, while for ACN, it
jumps up at the periphery. \fig{stressplots}a also shows that forces
in HSNs are much larger than those in ACNs with a comparable
equilibrium shape. The stress distribution in the boundary of an ACN
depends on the lattice constant $\ell$ of the network, as shown in
\fig{stressplots}b. The contour forces are minimal
in the middle and increase towards the sides. At an
adhesion site, all network forces add up to the overall force being
transmitted to the substrate. For both passive and active networks, the force
which is exerted on an adhesion site follows a power law as $\tau_H$
($\tau$) is increased, see \fig{stressplots}c. For $\tau=\tau_H$, adhesion force
is much smaller in the HSN than in the ACN. For active and passive
networks of a comparable shape, however, e.g. $\tau_H=0.2$ and
$\tau=0.01$, we observe the opposite behavior.

Although on an absolute scale the variation is not very strong,
\fig{cable_contr_hom}c and \fig{stressplots}b both demonstrate that
for ACNs under large network tension, peripheral force varies along
the contour. In contrast, the tension-elasticity model (TEM), which is
very successful in explaining shape, predicts homogeneous force
$\lambda=r\sigma$ along the boundary. \fig{cable_contr_hom}c suggests
one reason which could explain this discrepancy. For ACN, the links
essentially telescope in under contraction and therefore their density
along the contour varies for strong curvature along the contour. This
suggests that in order to explain the spatially varying force in the
contour, one has to revisit the assumption of an isotropic surface
tension $\sigma$ creating a homogeneous force density along the contour.

As an alternative to the TEM, we now investigate another analytical
model, which incorporates the effect of varying link density, namely
the elastic catenary \cite{lock2007}. In the elastic catenary, the
pulling force on the elastic contour is not along the normal, but along the
direction perpendicular to the original contour, similar to the
situation in networks with square shape and square topology, compare
\fig{cartoon}b. Due to the linear elasticity in the contour,
the line density of links along the contour varies in proportion
to the contour tension. In dimensionless units, a
length element is expanded to length $1+T(s)$ in the presence of 
tangential contour tension $\vec{T}(s) = T(s) \vec{t}(s)$.
With an initial inward force per unit length $\sigma$, 
the effective force density is therefore $\sigma / (1+T(s))$.
In contrast to the TEM, we now assume not only a heterogeneous
force distribution, but also a vertical pulling direction. 
Therefore the pulling force is 
$\vec{F}(s) = (0, - \sigma / (1+T(s)) )$. The force balance
again reads
\begin{equation}
\frac{d\vec{T}}{ds}+\vec{F}=0\ ,
\label{catgl_allg}
\end{equation}
but now the solution is more difficult than for the TEM.
Because the tangent is normalized, it can be written as $\vec{t} =
(\cos \phi(s), \sin \phi(s))$, where $\phi(s)$ is the tangential angle
along the contour \cite{whewell}, compare 
\fig{cartoon}b. Different from the TEM,
we now have to solve two equations:
\begin{eqnarray}
\frac{d}{ds} (T(s) \cos(\phi(s))) &=& 0\label{cat_tenx}\\
\frac{d}{ds} (T(s) \sin(\phi(s))) &=& \frac{\sigma}{1+T(s)}\ .\label{cat_teny}
\end{eqnarray}
\eq{cat_tenx} can directly be integrated,
leading to $T\cos(\phi)=const=\lambda_c$,  while \eq{cat_teny} can then be solved
via the substitution $\tan(\phi)=\sinh(p)$. This gives
\begin{eqnarray}
x(p)&=&\frac{\lambda_c}{\sigma}p+\frac{\lambda_c^2}{\sigma}
\sinh(p)+x_0,\\
y(p)&=&\frac{\lambda_c}{\sigma}\cosh(p)+\frac{\lambda_c^2}{2\sigma}
\cosh^2(p)+y_0.
\end{eqnarray}
The integration constants $x_0$ and $y_0$ are determined by the
positions of the adhesion sites. 

While in the TEM we assume the line tension $\lambda$ to be constant
along the whole boundary line, for the elastic catenary only the
$x$-component of tension, $\lambda_c$, is constant. Its value can be
determined numerically from the above equations for given $d$ and
$\sigma$. One can show that for small surface tension $\sigma$, the
contour becomes parabolic with a radius of curvature
$r=\lambda_c(1+\lambda_c)/\sigma$.  For $\sigma=10^{-2}$ the relative
deviation to the prediction of the TEM is only $4\%$, that is in this
regime, the elastic catenary model leads essentially to the same
result as the TEM with circular arcs. However, in contrast to the TEM,
this model predicts a spatially varying boundary tension of
\begin{equation}
T(p)=\lambda_c\cosh(p)\label{ten_c}\ .
\end{equation}
The curve without symbols in \fig{stressplots}b shows that
this model qualitatively predicts the observed minimum in the
stress distribution. The force acting on an adhesion dot is predicted to be
\begin{equation}
f=\sqrt{2}\lambda_c\left(1+\sinh(p_0)\right).\label{f_c}
\end{equation}
\fig{stressplots}d shows that this prediction is quite accurate.

\subsection{Strain Stiffening}

So far, we have treated the network links as cables or springs, in
which force increases linearly with elongation. This implies that
the elastic modulus of the network is independent of strain. 
However, it is well known that the cytoskeleton of cells 
shows a strong increase of elastic modulus with strain
(\textit{strain stiffening}) \cite{Janmey}. This property
of cytoskeletal networks has been reconstituted \textit{in vitro}
with crosslinked actin filaments of physiological length,
which is of the order of their persistence length (around 1 $\mu m$)
\cite{Gardel}. Under these conditions, strain stiffening
results mainly from the mechanical properties of
the crosslinkers. For larger filament lengths and 
stiff crosslinkers, there exists another mechanism for
strain stiffening, namely the non-linear force-extension
curve of semiflexible polymers as described by the
worm-like-chain (WLC) model \cite{doi}. The WLC-model has
been used before to model semiflexible biopolymers like
DNA~\cite{markosiggia}, actin~\cite{FreyKroy,FreyHeussinger} and
spectrin~\cite{sureshbpj,sureshpnas}. Because our network
model is especially suited to study the effect of link mechanics,
we now study strain stiffening based on the WLC-model, although
for the physiological relevant case, a model for crosslinker
mechanics might be more relevant.

\begin{figure}
\begin{center}
\includegraphics[width=0.8\columnwidth]{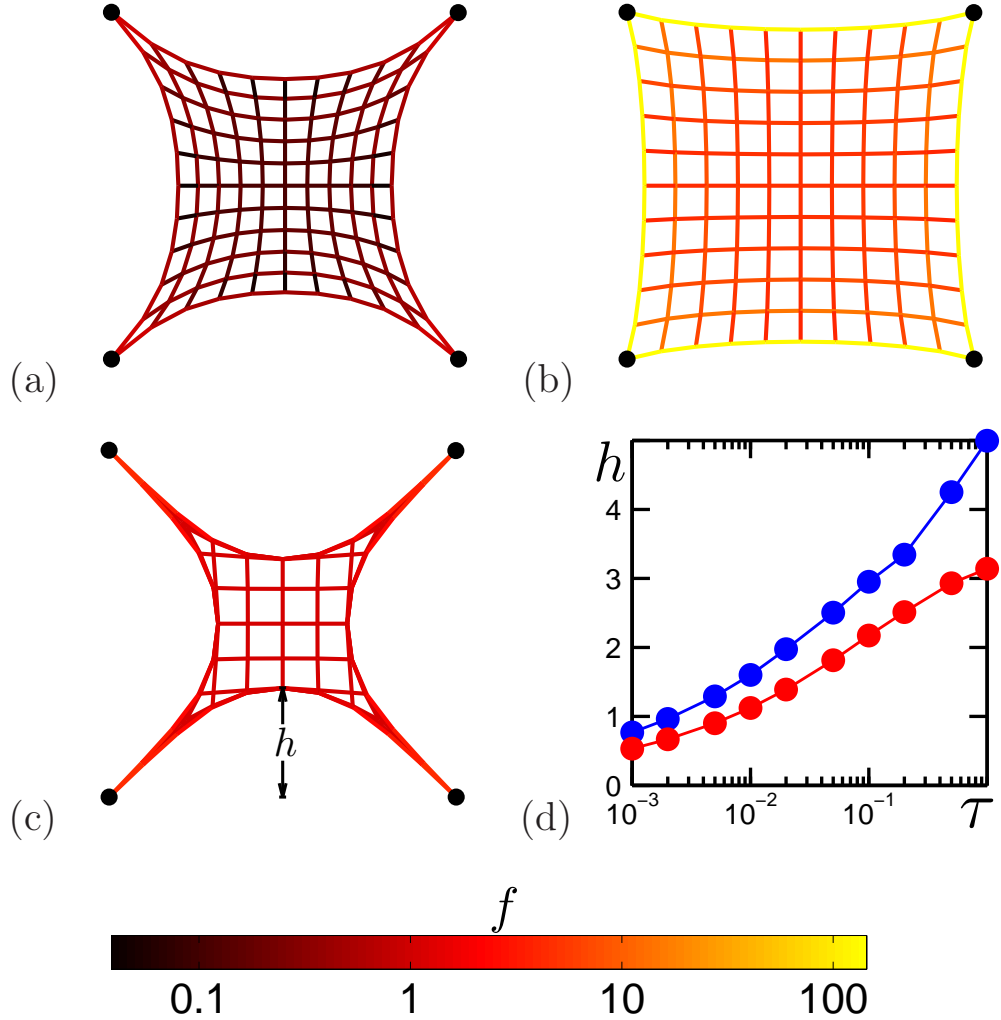}
\caption{(Color online) Contracted networks with worm-like chain (WLC) mechanics. (a) PCN with
 $\tau_H=0.47$. (b) PCN-WLC with $\tau_H=0.47$ and $\ell_n=0.5$.
 (c) ACN-WLC with $\tau=1$ and $\ell_n=0.5$. (d) Maximum invagination $h$ of
 linear (upper curve) and non-linear (lower curve) ACN.}
\label{wlc_pass_act}
\end{center}
\end{figure} 

While the WLC proper has vanishing resting length, here we combine it with a finite resting
length to also include the effect of compression. Thus we use the non-linear WLC
model to describe the mechanical links as they are tensed away from
their reference state, while the compressed state is modeled as above
(linear response for springs and no response for cables).
Complementing Eqs.\ (\ref{pass_cables1}) and (\ref{pass_cables2})
for the PCN, we get in dimensionless form
\begin{eqnarray}
\label{pass_wlc1}
\vec{f}_{ij}&=&\left({u}_{ij}+\frac{1}{4}
  \left(\frac{1}{(1-{u}_{ij})^2}
  -1\right)\right)\vec e_{ij} \qquad{1}<\ell_{ij}, \\
\vec{f}_{i j} &=& 0 \qquad{{\ell}_{ij}\leq1}.
\label{pass_wlc2}
\end{eqnarray}
Force is now given as multiples of $k_BT/L_0$, while length is again
scaled with $L_0$. Strain is now defined as ${u}_{ij}=(\ell_{ij}-1)/\ell_n$,
in which $\ell_n$ gives the dimensionless difference between
maximal extension and reference length. Effectively there is only one
difference to the original model, namely the
additional term which diverges if the strain ${u}_{ij}$
approaches $1$. Without strain, this term vanishes.
For ACNs, the original model is extended in the same way.

\begin{figure}
\begin{center}
\includegraphics[width=0.8\columnwidth]{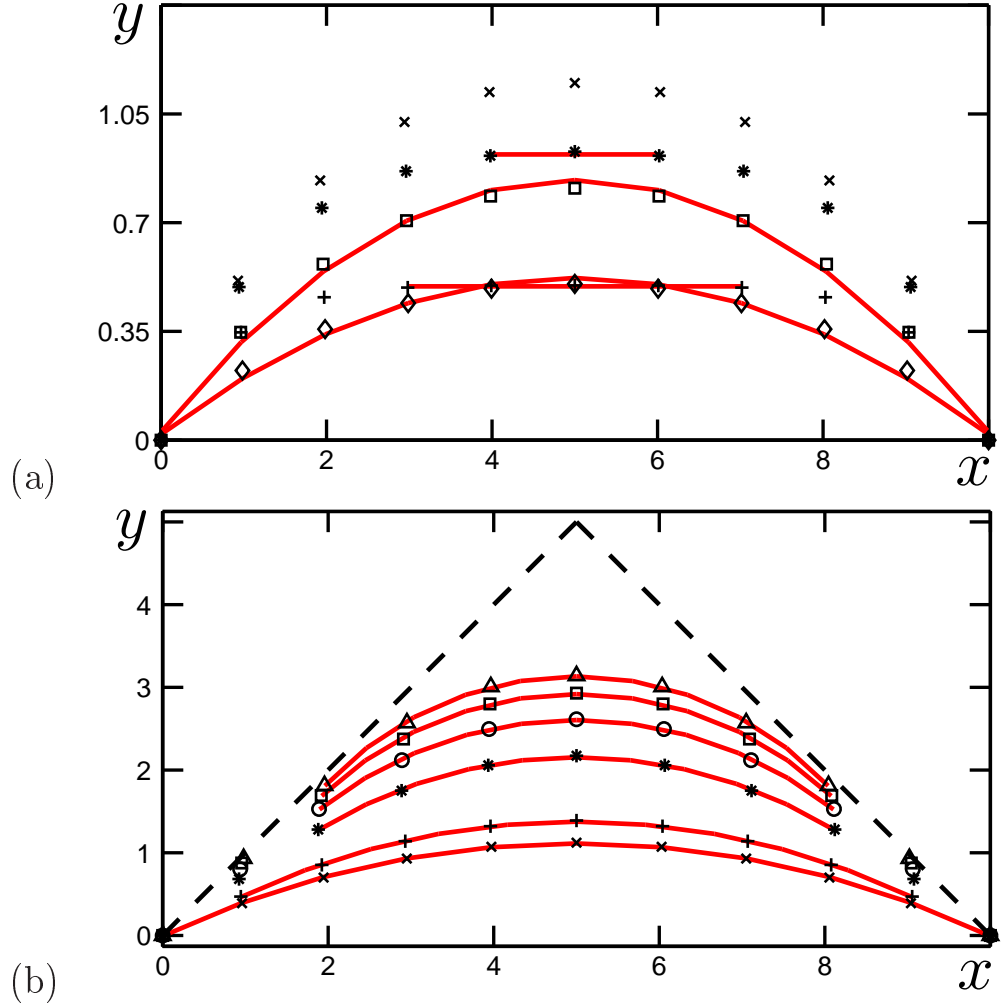}
\caption{(Color online) Arc fits for HSN and ACN with non-linear
 links. (a) Contour of the HSN-WLC from \fig{wlc_pass_act}b.
 Symbols belong to different values of $\tau_H$: $0.1$ ($+$), $0.2$
 ($*$), $0.3$ ($\times$), $0.43$ ($\square$), $0.47$ ($\Diamond$).
 $\Diamond$ correspond to the bottom line of the network from
 \fig{wlc_pass_act}b. (b) Contour of the ACN-WLC from
 \fig{wlc_pass_act}c. Symbols are bottom line node positions, while
 the lines are circular fits. Motor
 force values are $\tau=10^{-2}$ ($\times$), $2\cdot10^{-2}$ ($+$),
 $10^{-1}$ ($*$), $2.5\cdot10^{-1}$ ($\bigcirc$), $5\cdot10^{-1}$
 ($\square$), $1$ ($\bigtriangleup$). Note, $\bigtriangleup$ gives the
 bottom line from \fig{wlc_pass_act}c.}
\label{arcs_wlc}
\end{center}
\end{figure}

We choose $\ell_n=0.5$, that is the maximal extension is $1.5\ L_0$.
For small tension, $\tau_H<0.2$, the non-linearity does not affect the
shape of the PCN much and we observe the same invagination as
in \fig{hook_pass_contr}. \fig{wlc_pass_act}a,b shows for the PCN a comparison
between linear and WLC-networks for a large value of
tension, $\tau_H=0.47$. Obviously the strain-stiffened
network shows a much larger resistance to invagination. 
We also note that forces are two orders of magnitude larger in
the non-linear model.
ACNs are affected less by the non-linearity, as shown in
\fig{wlc_pass_act}c. Here we use $\tau=1$ and again
$\ell_n=0.5$. Comparison with \fig{cable_contr_hom}d reveals that the
ACN collapses to a lesser degree than without strain stiffening. This
can be quantified by the arc height $h(\tau)$, defined as the maximum
distance between initial and current edge in the equilibrium
shape. $h$ is significantly reduced by the non-linearity,
\fig{wlc_pass_act}d. Thus much higher motor forces are needed to reach
the collapsed state.

In \fig{arcs_wlc}a, the contour of the strain stiffening PCN is
analyzed in more detail. Circular and linear fits are shown as lines. The
bottom two are given by straight lines.  For $\tau_H<0.3$ the contour
of the PCN-WLC is qualitatively the same as that of the linear one,
shown in \fig{cable_arcs}b. At $\tau_H=0.3$ the contour cannot be
fitted well by circle or line.  If $\tau_H$ is increased beyond $0.3$,
the network does not contract any further, but again expands outward.
This surprising effect does not occur for ACNs. For
$\tau_H>0.4$, the arcs appear to be circular.  The ACN-WLC contour,
\fig{arcs_wlc}b, only differs little from the linear ACN contour,
\fig{cable_arcs}b. Arcs are always circular (except at the regions
where tubes form). With increasing $\tau$ they continuously move
inward. Comparison with \fig{cable_arcs}a reveals that radii typically
are larger in the non-linear case.

\subsection{Link Adaption}

\begin{figure}
\begin{center}
\includegraphics[width=0.8\columnwidth]{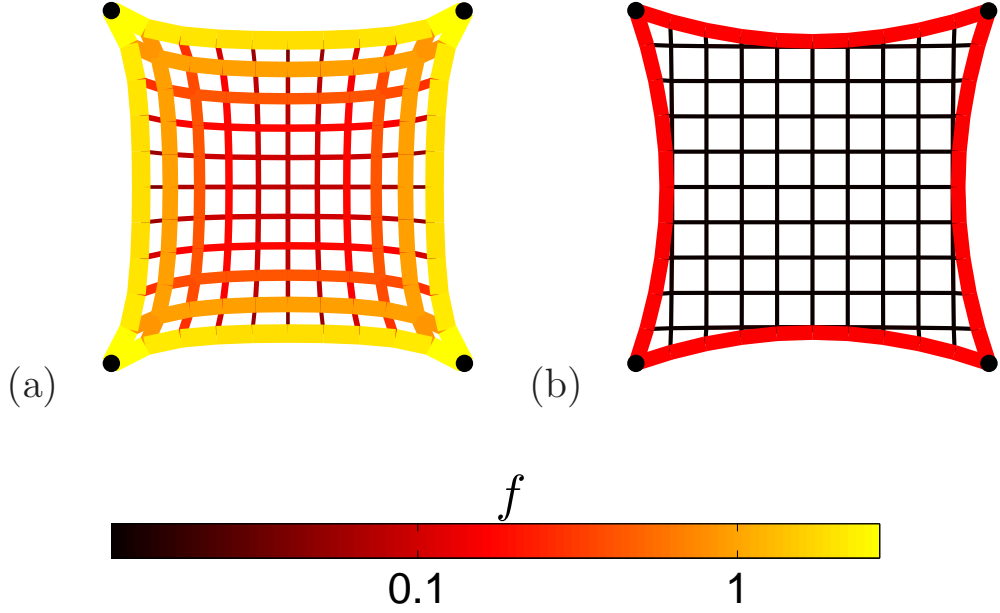}
\caption{(Color online) Equilibrium shapes for adaptive
 networks. (a) Square HSN/PCN network with $ea\approx9.40$ for
 the peripheral links and $ea\approx1.75$ for the innermost
 links. (b) Square ACN network with $ea\approx7.10$ at the periphery
 and $ea\approx2.00$ for the internal links. Parameters are:
 $[ea]_1=10$, $f_0=0.5$ (in (a)), $\tau_0=\tau_1=10^{-2}$,
 $[ea]_1=10$, $f_0=0.1$ (in (b)).}
\label{adap_states}
\end{center}
\end{figure}

Adherent cells are known to strongly adapt their cytoskeleton to the
physical properties of their environment. During recent years, it has
become clear that the actin cytoskeleton tends to reinforce under
load. In addition, mechanical loading of adhesion contacts leads to
regulatory signals which increase myosin motor activity inside the
cell. Network models are especially suited to study these biologically
important effects in a theoretical framework. In the following, we
will investigate which changes occur in the network if the elastic
constant $EA$ and the force density $T$ resulting from myosin II
activity are increasing with load.

For simplicity, we assume that both $EA$ and $T$ first increase with force
in a linear fashion and then saturate at constant values, which is the simplest
assumption for a process based on enzymatic regulation:
\begin{eqnarray}
EA(F)&=&[EA]_0+[EA]_1\frac{F}{F+F_0},\label{ea_f}\\
T(F)&=&T_0+T_1\frac{F}{F+F_0},\label{t_f}
\end{eqnarray}
where the force scale $F_0$ determines when half the maximal increase
has been reached.  We again use dimensionless parameters. Forces $EA$,
$T$, $F$ are measured in units $[EA]_0$, i.e.\ we define
$\tau=TL_0/[EA]_0$ (the same for $\tau_0$ and $\tau_1$), $ea=EA/[EA]_0$
(the same for $[ea]_1$, $[ea]_0=1$) and $f=F/[EA]_0$ (the same for
$f_0$).

\fig{adap_states} demonstrates that the effect of adaptation 
is fundamentally different for passive versus active networks.
Here stiffer links are represented by thicker lines. 
\fig{adap_states}a shows the result for a HSN (as shown by
\fig{hook_pass_contr}, for the square shape a PCN gives
the same results). In this case, only \eq{ea_f} must be considered.
While the peripheral links show the largest values of $ea$,
the rigidity decays smoothly from there into the bulk.
This is strikingly different for the ACN shown in 
\fig{adap_states}b. Here Eqs.\ (\ref{ea_f}) and
(\ref{t_f}) have been used. We find that the stiffness
is strongly localized to the periphery, as found before
for the internal stress. Thus for passive networks
the adaption response is spatially
continuous, while for actively contracting networks, it is strongly
localized to the rim. This nicely agrees with experimental observations
that strong peripheral actin bundles typically line the cell
contour \cite{bischofsbpj08}. In particular, our model suggests that
this effect is strongly determined by the mechanical properties
of the underlying networks. 

\subsection{Relation to Tissue Shape}

\begin{figure}
\begin{center}
\includegraphics[width=0.6\columnwidth]{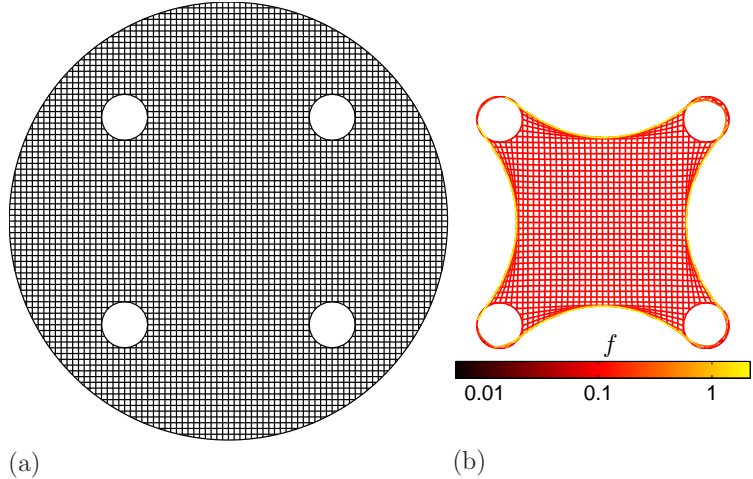}
\caption{(Color online) Contraction of an ACN anchored to dots of
 finite size. (a) Initial situation. Round tissue with radius $r_t$
 adherent to 4 round dots of finite radius $r_d$ which form a square
 with side length $d_d$: (b) Contracted tissue. Parameters are:
 $r_t=38$, $r_d=4$, $d_d=36$, $\tau=0.1$.}
\label{fat_dots}
\end{center}
\end{figure}

Tissue contraction with discrete pinning sites is very similar to cell
contraction since adherent tissues also show invaginated arcs
\cite{bischofsbpj08,chenpnas09}.  However, in this case the spatial
dimensions of the pinning objects tend to be relatively large.  In
order to include this effect, we have simulated a circular network
contracting around four circular dots of finite size, see
\fig{fat_dots}a. All nodes on the circles are fixed in space. Motor
force $\tau$ is increased stepwise and nodes coming closer to each
other than $\ell_c=0.01$ are glued together and in the following act
as one \cite{paul}. Relative dimensions are taken from
\cite{chenpnas09}, where a microtissue tethered to four cylindrical
posts is analyzed. With the ACN we are able to reproduce the typical
arc morphology of the contracted microtissues, see \fig{fat_dots}b.
We note that this is not possible with a FEM approach, as this leads
to flat contours as for HSN \cite{chenpnas09}. Thus ACN are
a useful model both for cells and tissues.

\section{Conclusions and Outlook}

Motivated by the network nature of the actin cytoskeleton and its
effectively 2D organization in mature adhesion to flat substrates, we
have modeled adherent tissue cell contraction by 2D network
models. The main aim of this work is to achieve a detailed comparison
of the shapes and force patterns for different network types, namely
Hookean spring networks (HSNs), passive cable networks (PCNs), and
active cable networks (ACNs).

The shape of a HSN can be understood best by considering the shape of
its unconstrained reference shape. If tension is not too large, the
network contour follows the unconstrained reference shape at regions
sufficiently far away from the adhesion sites.  Closer to the adhesion
sites, the network deforms and stress and strain accumulate.  In
contrast, the ACN does not have an unconstrained reference shape and
without adhesion constraints would contract into a point.  Therefore
no signature of the unconstrained reference shape (like flat parts for
a square-shaped lattice) appear in the contour.  Because it does not
resist compression, stress and strain are not propagated much into the
network and are strongly localized to the contour.

One of the most striking difference between the different network
types revealed by our analysis is the fact that in passive networks,
local changes to the adhesion geometry changes the network globally.
This is in marked contrast to the active network, where the addition
of local adhesions has only a local effect on the boundary. However,
in this case the change in spanning distance has a large effect on the
stress in the contour, as predicted by the tension-elasticity model
(TEM). The TEM is especially suited to quantitatively predict the
shape of an ACN, namely the circular arcs observed between
neighboring adhesion points and the scaling of their radius
with spanning distance and surface tension.

Despite this success, the TEM does not capture all aspects of the
network model. While the TEM assumes constant contour tension, the
computer simulations reveal that tension varies along the contour.
An elastic catenary theory qualitatively predicts that tension decreases
towards the middle of an invagination due to local changes in link
density along the contour. However, it does neither predict the
quantitative details of the contour stress nor the circular
arc morphology.

In the case of very large tension, both network types develop
different features. For the passive networks, the invaginations tend
to become more round, as the unconstrained reference shape becomes so
small that the contour cannot reach it anymore. In contrast, the
active network develops straight features, because the network
collapses into tubes at the adhesion points. Indeed the formation of
tubes has been observed experimentally and eventually leads to
pearling through a Rayleigh-Plateau instability \cite{c:barz99}. The
region between the tubes always stays circular for ACN.

Network models are ideally suited for multi-scale modeling because
physical properties can be easily added on the level of single link
and lead to non-trivial effects on the level of cell shape and forces.
In order to demonstrate this important aspect, we have studied two
important additional features of the cytoskeleton. First non-linear
links were introduced via the worm-like-chain model (WLC). In the WLC
case the passive square network first contracts and then expands again
as tension is increased. The ACN requires much larger values of
tension to contract compared to the linear case. Otherwise the arc
morphology is essentially the same as in the linear case.

As a biologically very relevant aspect of the cellular cytoskeleton,
we also have studied the adaptation response of network links.
For certain parameter values, a saturation response for both
elasticity and tension leads to a strong difference between $EA$ and
$T$ of boundary links, which are strongly increased, and $EA$ and $T$
of internal links, which are increased much less. Similar aspect have
been addressed before in the framework of Finite Element Modeling
(FEM) \cite{EvansPNAS}. In this case, the biochemical regulation has
been modeled with more detail. Both the resulting cell shapes and the
formation of stress fibers inside the cell demonstrate that the
FEM model strongly resembles the HSN studied here. Therefore
it would be interesting to combine the detailed biochemical model
with the actively contracting cable network studied here.
While it appears to be very challenging to develop a homogenization
strategy for the ACN, it is interesting to consider if similar
features as resulting from the ACN could be obtained in a FEM-framework.

Active cable networks have also been shown to describe the circular
arc morphology of tissues pinned at discrete sites
\cite{bischofsbpj08}.  Because here arc radius also scales with
spanning distance as for the arc radius of strongly adhering cells,
the tension-elasticity model seems to capture all essential element of
this situation.  In the tissue case, the cable network represents the
fibrous nature of the collagen matrix and the active contractility
corresponds to cell contraction. Because in addition water can flow
out of the contracting cell-matrix composite, volume is not conserved
and compression is not propagated. Therefore the standard models of
elasticity are strictly speaking not appropriate. Indeed they do not
predict the circular arc morphology, but rather show flat contours
corresponding to the unconstrained reference shape of the elastic
model \cite{chenpnas09}.

In summary, HSN, PCN and ACN are simple model systems which however
show surprisingly rich responses to internal contractility and
therefore lead to interesting conclusions about the physical elements
required to endow cells with a sense of geometry. ACN seem to be very
appropriate to model strongly adherent tissue cells as they not only
implement some of the most important fundamental features of the
cytoskeleton (asymmetry under tension and compression, contraction by
molecular motors), but also lead to functions which are very reminiscent
of real cells (robustness under structural re-arrangements and
adaptation to local adhesion constraints).

\begin{acknowledgments}
USS is a member of the Heidelberg cluster of excellence
CellNetworks and was supported by the MechanoSys-grant
from the Federal Ministry of Education and Research (BMBF) of
Germany. IBB was supported by the German Science Foundation (DFG)
through the Emmy Noether program (BI1213-3/1).
\end{acknowledgments}


\end{document}